\documentclass[aps,preprint,tightenlines,nofootinbib,amssymb,superscriptaddress]{revtex4}
\usepackage{amssymb}
\usepackage{mathrsfs}
\usepackage{slashed}
\usepackage{graphicx,color}
\usepackage{amsmath}
\allowdisplaybreaks

\begin{document}
\title{LHC Accessible Second Higgs Boson in the Left-Right Model}
\author{Rabindra N. Mohapatra}
\email{rmohapat@umd.edu}
\affiliation{Maryland Center for Fundamental Physics and Department of Physics, University of Maryland, College Park, Maryland 20742, USA}
\author{Yongchao Zhang}
\email{yongchao@umd.edu}
\affiliation{Center for High Energy Physics, Peking University, Beijing, 100871, P. R. China}
\affiliation{Maryland Center for Fundamental Physics and Department of Physics, University of Maryland, College Park, Maryland 20742, USA}
\date{\today}
\hfill{UMD-PP-014-001}

\begin{abstract}
  A second Higgs doublet  arises naturally as a parity partner of the standard model (SM) Higgs, once SM is extended to its left-right symmetric version (LRSM) to understand the origin of parity violation in weak interactions as well as to accommodate small neutrino masses via the seesaw mechanism. The flavor changing neutral Higgs (FCNH) effects in the minimal version of this model (LRSM), however, push the second Higgs mass to more than 15 TeV making it inaccessible at the LHC. Furthermore since the second Higgs mass is directly linked to the $W_R$ mass, discovery of a  ``low'' mass  $W_R$ ($M_{W_R}\leq 5-6$ TeV) at the LHC would require values for some Higgs self couplings larger than one.  In this paper we present an extension of LRSM by adding a vector-like $SU(2)_R$ quark doublet which weakens the FCNH constraints allowing the second Higgs mass to be near or below TeV and a third neutral Higgs below 3 TeV for a $W_R$ mass below 5 TeV. It is then possible to search for these heavier Higgs bosons  at the LHC, without conflicting with FCNH constraints. A right handed $W_R$ mass in the few TeV range is quite natural in this class of models without having to resort to large scalar coupling parameters. The CKM mixings are intimately linked to the vector-like quark mixings with the known quarks, which is the main reason why the constraints on the second Higgs mass is relaxed. We present a detailed theoretical and phenomenological analysis of this extended LR model and point out some tests as well as its potential for discovery of a second Higgs at the LHC.  Two additional features of the model are: (i) a 5/3 charged quark and (ii) a fermionic top partner with masses in the TeV range.
\end{abstract}
\maketitle

\tableofcontents

\newpage
\section{Introduction}
With the discovery of the standard model Higgs boson at the Large Hadron Collider~\cite{lhchiggs}, attention has now shifted in part to the search for a second heavier Higgs boson~\cite{2higgs,2hth}. While the 125 GeV Higgs boson has confirmed the standard model, the second Higgs field is likely to provide strong clues to the nature of new physics beyond the standard model. For instance, a second Higgs boson is a natural part of several extensions of SM e.g. minimal supersymmetric standard model (MSSM), Peccei-Quinn extension to solve the strong CP problem as well as models with spontaneous CP violation as in  multi-Higgs extensions of SM\cite{branco}. Another class of models where also a second Higgs doublet is forced on us by gauge symmetry is the left-right symmetric (LRSM) extension of SM \cite{LR}, which provides a way\cite{LRseesaw} to understand the small neutrino masses via the seesaw mechanism. The second Higgs doublet in this model is the parity partner of the SM Higgs and is dictated by  the gauge group of the LRSM, $SU(2)_L\times SU(2)_R\times U(1)_{B-L}$. The minimal version of this model (MLRSM) is defined as the one with parity symmetric fermion and Higgs assignments with scalar bidoublet field $\phi(2,2,0)$ giving masses to charged fermions and the $\Delta_L(3,1,2)\oplus \Delta_R(1,3,2)$ breaking the $SU(2)_R\times U(1)_{B-L}$ symmetry as well as implementing the seesaw mechanism. In this minimal version, the SM Higgs  field ($\phi_{SM} $) is part of a bi-doublet field $\phi$ which contains two SM doublets, the second being the parity partner of  $\phi_{SM} $. It turns out however that in MLRSM, gauge invariance also restricts the coupling of the second Higgs to the quarks in such a way that it leads to large flavor changing neutral Higgs effects unless its mass is more than about 15 TeV~\cite{Zhang:2007da}. This pushes the second Higgs boson beyond the reach of LHC but perhaps more importantly, if a right handed $W_R$ is discovered at the LHC with mass below 5-6 TeV, some self scalar coupling parameter in the potential must be larger than one, causing some tension.

This also raises the following more practical question: suppose LHC discovers a second Higgs boson with a few TeV (or less) mass; in that case, should the search for $W_R$ boson at the LHC~\cite{CMSWR} stop ? The minimal LRMS would say ``yes" since the second TeV-ish Higgs boson would rule this model out. What we point out in this paper is that  a second TeV-ish neutral Higgs boson does not necessarily rule out the general class of $SU(2)_L\times SU(2)_R\times U(1)_{B-L}$ extensions of SM since as we show in this paper, there exist non-minimal left-right models with few TeV mass $W_R$ which can accommodate a near TeV mass Higgs boson without conflicting with FCNH constraints from meson-anti-meson mixings.  The search for TeV scale $W_R$ boson at the 14 TeV LHC should therefore continue even if a TeV-ish Higgs boson is discovered. In fact we point out that in the example we propose, the search for $W_R$ should most likely continue in the tri-lepton mode rather than the $\ell^\pm\ell^\pm jj$ mode~\cite{ks,gin} currently being used.  While this may not be a generic feature of such extended models, it may be something to keep in mind until a different example is found. The goal of this paper is to provide an existence proof by example of such a model\footnote{For an alternative approach, see~\cite{yueliang}.}.

It was pointed few years ago that if we imagine the TeV scale left-right model as an effective theory, then one can add higher dimensional ($d=6$) operators to the theory that involve additional Higgs fields and the minimal LRSM fermions which can help to bring consistency between the TeV scale Higgs mass with FCNH effects~\cite{guada}. The example provided in this paper is a UV complete version of the model which requires adding a single vector-like $SU(2)_R$ quark doublet and an extra Higgs bi-doublet carrying $B-L$. We show that this model reproduces the CKM mixing in the quark sector with tiny deviations from unitarity and we then analyze the FCNH effects which exist only in the up-quark sector, i.e. in the $D^0-\bar{D^0}$ mixing. We find that for the second Higgs boson mass near a TeV, we can satisfy the FCNH bounds. There are no FCNH effects in the down quark sector due to the symmetries of the model. As part of our study of phenomenological implications of the model, we discuss production and decay modes of the second heavy Higgs.

The basic outline of our strategy is as follows. The root of the FCNH bound on the second Higgs in the minimal LRSM is that the bi-doublet field $\phi$ and its conjugate $\tilde{\phi}$ both couple to quark doublets as ${\cal L}_Y~=~h\bar{Q}_L\phi Q_R+\tilde{h}\bar{Q}_L\tilde{\phi}Q_R~+~h.c.$. When quark mass matrices are diagonalized to generate the $V_{CKM}$, the neutral component of the second Higgs doublet has off diagonal couplings which then give rise to the Flavor changing effects. The first step to cure this problem is the prevent  the $\tilde{\phi}$ Yukawa coupling to quarks, while allowing the $\phi$ coupling. This however leads to $V_{CKM}=1$ so that all quark mixings vanish. Our suggestion to cure this problem is to introduce one set of vector like quarks which are such that generate the quark mixings only in the up sector. As a result, the only FCNH effect we have to consider is the $D^0-\bar{D^0}$ mixing. Since CKM mixings arise due to small mixings with the vector like quarks, the resulting constraint on the second neutral Higgs is much weaker. The presence of the extra bi-doublet also leads to a third Higgs field with mass below 3 TeV for a $W_R$ mass less than 5 TeV. While we do not go into details of the lepton sector of our model, we note that small neutrino masses most likely arise in this model from the inverse seesaw mechanism~\cite{inverse}. This in turn implies that the search for $W_R$ should focus on the tri-lepton mode~\cite{dev1}.

The paper is organized as follows: In sec. 2, we present the particle content and the fermion sector of the model and then obtain a parameter range where the correct CKM mixings arise; in sec. 3, we discuss the Higgs potential, its minimization to obtain the neutral Higgs spectrum; in sec. 4, we find the FCNH constraints on the Higgs masses. In sec. 5, we choose some bench mark points of the model and discuss the LHC signal for the Higgs fields of the model. In sec. 6, we briefly touch on the lepton sector of the model, more specifically, the origin of neutrino masses. In sec. 7, we discuss some other phenomenological implications as well as comment on the grand unification prospects for the model.  We present a summary of our results in the final section 8. In appendix A, we present an example of quark mixing solution without CP violation and in appendix B, we give details of the potential minimization and neutral scalar mass diagonalization.

\section{Extended left-right model}

In addition to the usual left-right fermion doublets  $Q^T_{a,L,R}= (u_a, d_a)_{L,R}$ and $\ell^T_{a, L,R} =(\nu_a, e_a)_{L,R}$ with obvious $U(1)_{B-L}$ quantum numbers, we add an $SU(2)_R$ vector like quark doublet $Q^{\prime T}\equiv (T, t')$ with $B-L=\frac73$. Because of the exotic B-L assignment, the vector-like quark $T$ has electric charge 5/3 and $t'$ has $Q=2/3$ like the top quark. We also note that the model at the TeV scale does not respect discrete parity invariance. However, adding extra heavy $SU(2)_L$ vector-like quarks and parity odd Higgs fields, the model can be made parity invariant at a high scale. As a result of this high scale breakdown of  parity there is no type II contribution to neutrino masses.

The Higgs sector of the extended LRSM model suggested here consists of the following Higgs fields, with their quantum numbers under the $SU(2)_{\rm L} \times SU(2)_{\rm R} \times U(1)_{\rm B-L}$ gauge symmetry given within bracket:
\begin{equation}
\begin{array}{ll}
\vspace{.2cm}
\phi = \left( \begin{array}{cc} \phi_{1}^0 & \phi_{2}^{+} \\
\phi_{1}^{-} & \phi_{2}^0\\ \end{array} \right) & \in (2,2,0) \,, \\ \vspace{.2cm}
\rho = \left( \begin{array}{cc} \rho_{1}^+ & \rho_{}^{++} \\
\rho_{}^{0} & \rho_{2}^+ \\ \end{array} \right) & \in (2,2,2) \,, \\
\vspace{.2cm}
\Delta_L = \left(
\begin{array}{cc}\delta_{L}^+ / \sqrt{2}
& \delta_{L}^{++} \\ \delta_{L}^{0} & -\delta_{L}^+ / \sqrt{2}\\
\end{array} \right) & \in (3,1,2) \,, \\
\Delta_R = \left( \begin{array}{cc} \delta_{R}^+ / \sqrt{2} &
\delta_{R}^{++} \\ \delta_{R}^{0} & -\delta_{R}^+ / \sqrt{2}\\
\end{array} \right) & \in (1,3,2) \,.
\end{array}
\end{equation}
Under the $SU(2)_{\rm L} \times SU(2)_{\rm R}$ gauge symmetry, these fields transforms as
\begin{equation}
\begin{array}{ll}
\vspace{.1cm}
\phi             \rightarrow U_L \phi             U_R^\dagger \,, &
\tilde\phi       \rightarrow U_L \tilde\phi       U_R^\dagger \,, \\
\vspace{.1cm}
\rho             \rightarrow U_L \rho             U_R^\dagger \,, &
\tilde\rho       \rightarrow U_L \tilde\rho       U_R^\dagger \,, \\
\vspace{.1cm}
\Delta_L         \rightarrow U_L \Delta_L         U_L^\dagger \,, &
\Delta_L^\dagger \rightarrow U_L \Delta_L^\dagger U_L^\dagger \,, \\
\Delta_R         \rightarrow U_R \Delta_R         U_R^\dagger \,, &
\Delta_R^\dagger \rightarrow U_R \Delta_R^\dagger U_R^\dagger \,.
\end{array}
\end{equation}
where $\tilde\varphi = -i\sigma_2 \varphi^\ast i\sigma_2$ ($\varphi = \phi,\,\rho$), and $U_{L,\,R}$ are, respectively, the general $SU(2)_{\rm L}$ and $SU(2)_{\rm R}$ unitarity transformations.

We assume the theory to be invariant under a discrete $Z_4$ symmetry so that simultaneous coupling of $\phi$ and $\tilde{\phi}$ couplings to the SM quarks is forbidden naturally. The complete set of transformations of all the fields in the extended LRSM  under the discrete $Z_4$ symmetry are given below:
\begin{equation}
\begin{array}{lll}
\vspace{.1cm}
\phi \rightarrow i \phi \,,  &
Q_L \rightarrow Q_L \,, &
\ell_L \rightarrow \ell_L \,, \\
\vspace{.1cm}
\rho \rightarrow i \rho \,, &
Q_R \rightarrow -i Q_R \,, &
\ell_R \rightarrow i \ell_R \,. \\
\Delta_R \rightarrow - \Delta_R \,, &
Q'_{L,\,R} \rightarrow i Q'_{L,\,R} \,,
\end{array}
\end{equation}
Consequently, for the two scalars $\phi$ and $\rho$,
\begin{eqnarray}
\tilde{\varphi} \rightarrow -i \tilde\varphi \,.
\end{eqnarray}

The Yukawa couplings of the model read, under the discrete symmetry,
\begin{eqnarray}
-\mathcal{L}_Y &=&
\bar{Q}_L h_q\phi Q_R  + \bar{Q}_L y_f \tilde\rho Q'_R
   + \bar{Q}'_L y_g \Delta_R Q_R
   + M \bar{Q}'_L Q'_R   \nonumber\\
  && +\bar{\ell}_L h_\ell \tilde{\phi} \ell_R + y_R \ell_R \ell_R \Delta_R ~+~{\rm h.c.} \,.
\end{eqnarray}
After the scalars get non-vanishing vevs (for simplicity we assume all the vevs are real),
\begin{eqnarray}
&& \langle \phi \rangle = \frac{1}{\sqrt2} \left( \begin{array}{cc}
\kappa_1 & 0 \\ 0 & \kappa_2 \\
\end{array} \right), \nonumber \\
&& \langle \rho \rangle = \frac{1}{\sqrt2} \left( \begin{array}{cc}
0 & 0 \\ v_\rho & 0 \\
\end{array} \right), \nonumber \\
&& \langle \Delta_R \rangle = \frac{1}{\sqrt2} \left(
\begin{array}{cc} 0 & 0 \\ v_R & 0\\
\end{array} \right) \ ,
\end{eqnarray}
the quark mass matrices read
\begin{eqnarray}
&& \mathcal{M}_d =\frac{1}{\sqrt{2}} {\kappa_2}h_q \,, \nonumber \\
&& \mathcal{M}_u =\frac{1}{\sqrt{2}}  \left( \begin{array}{cc}
h_q\kappa_1 & y_f v_\rho \\ y_g v_R & \sqrt{2}M \end{array} \right)
= \left( \begin{array}{cc}
h & f \\ g & M \end{array} \right) \,,
\end{eqnarray}
where $h_q$ is a $3\times3$ matrix which can be chosen to be diagonal by a choice of quark basis, and consequently
\begin{eqnarray}
&& \mathcal{M}_d = {\rm diag} \{ m_d,\, m_s,\, m_b \} \,.
\end{eqnarray}
Then the quark mixings as well as CP violation, i.e. a nontrivial CKM matrix, comes from the $f$ and $g$ parameters, or more specifically from the Yukawa couplings $y_f$ and $y_g$. Note that the $y_f$ and $y_q$ are not related by left symmetry since they are coupled to different Higgs bosons not related by parity at high scale. This means that the left and right handed quark mixing angles will be very different from each other.

\subsection{CKM fit in the model}
As we see from the previous sub-section, the quark mass matrices are very highly constrained and therefore it is  a priori not clear that the model will reproduce the correct quark masses and CKM mixings for reasonable choice of parameters. Below we show that this is indeed the case. The starting point of quark mixing is the $4\times4$ up-type mass matrix in the extended LRSM,
\begin{eqnarray}
\mathcal{M}_u = \left( \begin{matrix}
h_1 & 0 & 0 & f_1 \\
0 & h_2 & 0 & f_2 \\
0 & 0 & h_3 & f_3 \\
g_1 & g_2 & g_3 & M
\end{matrix} \right) \,.
\end{eqnarray}
This matrix $\mathcal{M}_u\mathcal{M}_u^\dagger$ can be diagonalized by the $4\times4$ CKM matrix $V_{\rm CKM4}$, in the basis with diagonal down-type quark mass matrix,
\begin{eqnarray}
V_{\rm CKM4} (\mathcal{M}_u\mathcal{M}_u^\dagger) V_{\rm CKM4}^\dagger = {\rm diag} \{ m_u^2,\, m_c^2,\, m_t^2,\, m^{\prime2}_t \} \,,
\end{eqnarray}
where $m'_t$ is the mass for the introduced sequential 4th up-type quark. Equivalently,
\begin{eqnarray}
\label{diagonalization}
\mathcal{M}_u\mathcal{M}_u^\dagger = V_{\rm CKM4}^\dagger {\rm diag} \{ m_u^2,\, m_c^2,\, m_t^2,\, m^{\prime2}_t \} V_{\rm CKM4} \,.
\end{eqnarray}

On the LHS of the equation, in our extended LRSM,
\begin{eqnarray}
&& h_1 = r m_d \,, \nonumber \\
&& h_2 = r m_s \,, \nonumber \\
&& h_3 = r m_b \,,
\end{eqnarray}
where $r$ is the vev ratio $\kappa_1/\kappa_2$. As the up-type quark mass matrix is diagonalized at the TeV scale, i.e. the new physics scale of our model, we use the RGE-evolved down-type and up-type quark masses in our fit~\cite{Xing:2007fb},
\begin{equation}
\begin{array}{ll}
\vspace{.1cm}
m_u ({\rm TeV}) = 2.6\,{\rm MeV} \,, &
m_d ({\rm TeV}) = 2.5\,{\rm MeV}\,, \\
\vspace{.1cm}
m_c ({\rm TeV}) = 0.53\,{\rm GeV} \,, &
m_s ({\rm TeV}) = 47 \,{\rm MeV}\,, \\
m_t ({\rm TeV}) = 150.7\,{\rm GeV} \,, &
m_b ({\rm TeV}) = 2.43\,{\rm GeV}\,.
\end{array}
\end{equation}
On the RHS of Eq.~(\ref{diagonalization}), the unitary of the $3\times3$ CKM matrix is in good agreement with observations, leaving little room for mixing with heavy quarks, i.e.~\cite{pdg2013}
\begin{eqnarray}
&& |V_{ud}|^2 + |V_{us}|^2 + |V_{ub}|^2 = 0.9999 \pm 0.0006 \,, \nonumber \\
&& |V_{cd}|^2 + |V_{cs}|^2 + |V_{cb}|^2 = 1.067 \pm 0.047 \,, \nonumber \\
&& |V_{td}|^2 + |V_{ts}|^2 + |V_{tb}|^2 = 1. \pm 0.000137 \,.
\end{eqnarray}

In Appendix A, We give a toy fit of the CKM matrix without any CP violation, which however reveal some specific feature of the fitting, e.g. $r$ is required to be of order 10, and the $f_i$ parameters are generally small while $g_i$ are TeV scale parameters. In the realistic CP violating case, $M$ is kept as a real parameter, while $f_j$ and $g_j$ are required to have non-vanishing phases denoted as
\begin{eqnarray}
&& f_j \rightarrow f_j e^{i\alpha_j} \,, \nonumber \\
&& g_j \rightarrow g_j e^{i\beta_j} \,,
\end{eqnarray}
then the LHS of Eq.~(\ref{diagonalization}) reads
\begin{eqnarray}
&& \mathcal{M}_u\mathcal{M}_u^\dagger = \nonumber \\
&& \left( \begin{matrix}
h^2_1 +f^2_1 & f_1f_2 e^{i(\alpha_1-\alpha_2)} & f_1f_3 e^{i(\alpha_1-\alpha_3)} & f_1M e^{i\alpha_1} + g_1h_1 e^{-i\beta_1} \\
f_1f_2 e^{-i(\alpha_1-\alpha_2)} & h^2_2 +f^2_2 & f_2f_3 e^{i(\alpha_2-\alpha_3)} & f_2M e^{i\alpha_2} + g_2h_2 e^{-i\beta_2} \\
f_1f_3 e^{-i(\alpha_1-\alpha_3)} & f_2f_3 e^{-i(\alpha_2-\alpha_3)} &
h^2_3 +f^2_3 & f_3M e^{i\alpha_3} + g_3h_3 e^{-i\beta_3} \\
f_1M e^{-i\alpha_1} + g_1h_1 e^{i\beta_1} &
f_2M e^{-i\alpha_2} + g_2h_2 e^{i\beta_2} &
f_3M e^{-i\alpha_3} + g_3h_3 e^{i\beta_3} &
g^2_1 +g^2_2 +g^2_3 +M^2
\end{matrix} \right) \,. \nonumber \\
\end{eqnarray}

One representative solution we find for the CP violating case is  the following:
\begin{eqnarray}
&& h_1 = 0.037 \,, \nonumber \\
&& h_2 = 0.346 \,, \nonumber \\
&& h_3 = 13.1 \,;
\end{eqnarray}
for the $f$ parameters (in unit of GeV) and its phases (in unit of radian)
\begin{equation}
\begin{array}{ll}
f_1 = 1.38 \,, & \alpha_1 = -2.80 \,, \\
f_2 = 6.32 \,, & \alpha_2 = -0.0499 \,, \\
f_3 = 158 \,, & \alpha_3 = -3.17 \,;
\end{array}
\end{equation}
for the $g$ parameters (in unit of GeV) and its phases (in unit of radian)
\begin{equation}
\begin{array}{ll}
g_1 = 1450 \,, & \beta_1 = 1.64 \,, \\
g_2 = 2398 \,, & \beta_2 = -2.93 \,, \\
g_3 = 573 \,, & \beta_3 = 2.43 \,;
\end{array}
\end{equation}
and the last parameter, in unit of GeV,
\begin{eqnarray}
\label{No.25}
M = 905 \,.
\end{eqnarray}
For the $h$ parameters, if we choose the mass ratio $r=5$, then the down-type quark masses we need as input are, respectively,
\begin{eqnarray}
&& m_d ({\rm TeV}) = 7.57 \,{\rm MeV}\,, \nonumber \\
&& m_s ({\rm TeV}) = 69.4 \,{\rm MeV}\,, \nonumber \\
&& m_b ({\rm TeV}) = 2.62\,{\rm GeV}\,.
\end{eqnarray}
which are consistent with their experimental values. With these parameters we can fit successfully all the up-type quark masses, left-handed quark mixing angles and CP violation phase. Furthermore, much like in the CP conserving case, the mixing between the SM up-type quarks with $t'$ are very small,
\begin{eqnarray}
&& s_{14} = 0.000025 \,, \nonumber \\
&& s_{24} = 0.00012 \,, \nonumber \\
&& s_{34} = 0.0165 \,.
\end{eqnarray}
As a direct consequence of the large $g_i$ parameters, however, contrary to the minimal LRSM case~\cite{Zhang:2007da}, the right-handed quark mixings are generally very large, e.g. in the fit above,
\begin{eqnarray}
\left( \begin{array}{cccc}
 0.142 +0.779 i & 0.513 +0.0444 i & -0.32744-0.0112 i & -0.0272 \\
 -0.0286+0.337 i & 0.101\, -0.128 i & 0.923 +0.0314 i & 0.0765 \\
 -0.0171+0.158 i & -0.255-0.0654 i & 0.0375\, +0.0425 i & -0.950 \\
 -0.0336+0.482 i & -0.781-0.169 i & -0.145+0.124 i & 0.302 \\
\end{array} \right) \,.
\end{eqnarray}
Due to the large $W_R$ mass, however, they do not lead to any conflict with observations.

\section{Spectrum of Higgs fields}
Now that the model can reproduce the quark masses and mixings, we move on to discuss the masses and decay properties of the second and other neutral Higgs in the model. Our interest is primarily in the second Higgs which consists predominantly of the components  Re$\phi^0_{1,\,2}$ and Re$\rho^0$. 
Our model has four complex neutral Higgs fields $\phi^0_{1,2}, \rho^0, \delta^0_R$ which will in general mix among themselves. We have to find the mass eigenstates. To proceed with this study, we analyze the gauge and $Z_4$ invariant Higgs potential Eq.~(\ref{potential}) in Appendix B allowing only for two soft $Z_4$ breaking terms (Eq.~(\ref{soft})). We then minimize the Higgs potential to obtain the desired minima given in Eq. (6) above and use them to find the spectrum of neutral Higgs fields, which are mass eigenstates~\cite{Deshpande:1990ip}.  Note that unlike the minimal LRSM, there is no $\Delta_L$ field. We adopt high scale D-parity breaking to push this field to very high scale~\cite{CMP}, which also eliminates the type II contributions to neutrino masses that could otherwise be ``large" in TeV LR models.

\begin{figure}[t]
  \begin{center}
  \includegraphics[width=7.5cm]{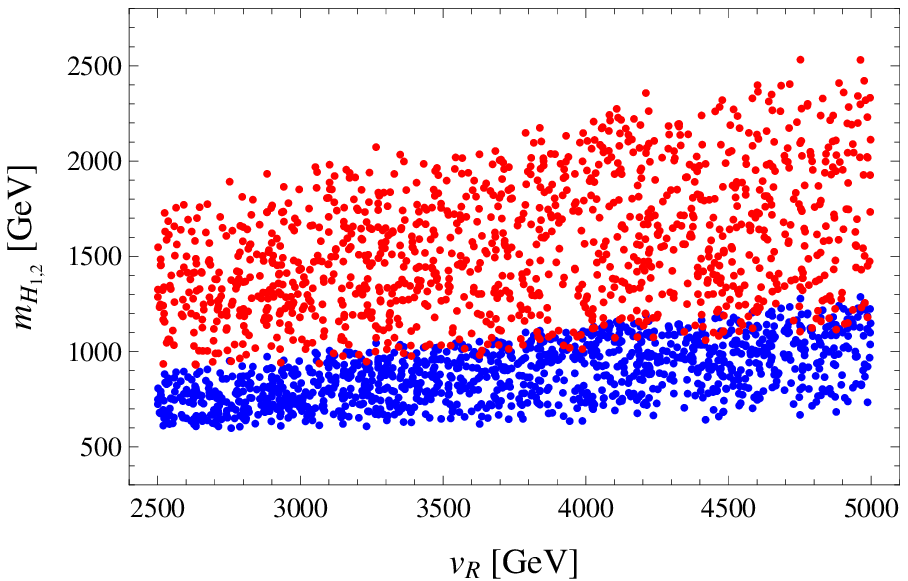}
  \includegraphics[width=7.5cm]{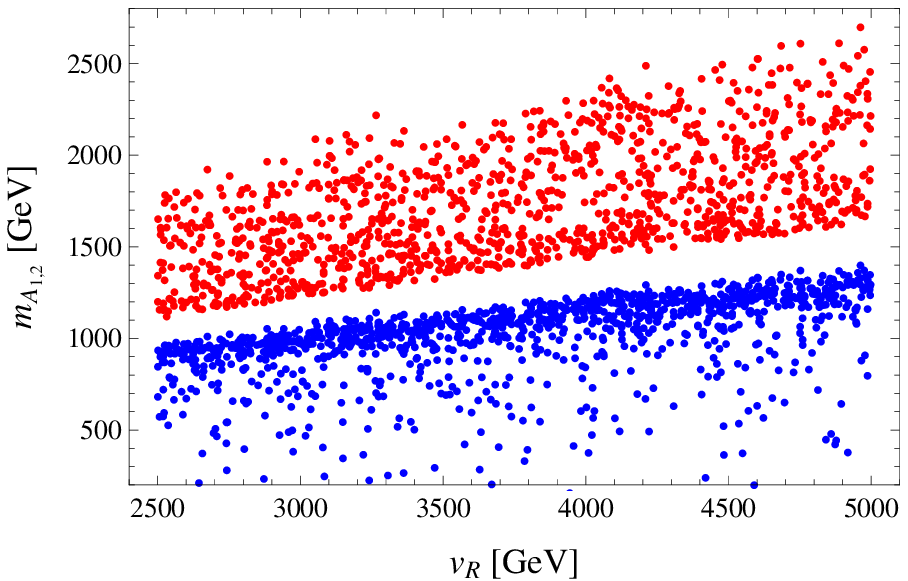}
  \vspace{-.3cm}
  \caption{The masses of $H_{1,\,2}$ and $A_{1,\,2}$ as function of $v_R$. The masses of $A_1$ and $H_1$ are in blue and $A_2$ and $H_2$ are in red.}
  \vspace{-.7cm}
  \label{scattering-plot}
  \end{center}
\end{figure}

For numerical fit, after setting a sum of two scalar couplings (denoted by $\gamma'$, cf. Eq.~(\ref{parameter-prime})) $\gamma'=0.5$, we  get a scalar $H_R$ with mass $v_R$ at the leading order. By assigning appropriate values to other relevant coefficients and with the $SU(2)_L$ breaking vevs satisfying the constraint\footnote{Numerical analysis reveals that only about one per cent of the SM higgs is from its mixing with the right-handed Higgs $\delta_R^0$. As a leading order approximation, we can neglect this contribution.}
\begin{eqnarray}
\kappa_1^2 + \kappa_2^2 + v_\rho^2 = v^2 = (246\,{\rm GeV})^2 \,,
\end{eqnarray}
we can easily get a 125 GeV Higgs recently observed at the LHC; we can also obtain the corresponding masses of the other two CP-even states, shown in Fig.~\ref{scattering-plot} as functions of $v_R$. It is obvious that $m_{H_1} \sim \sqrt{vv_R}$ and $H_2$ has a somewhat larger and broader mass range\footnote{In the numerical fit, we set explicitly $\gamma'=0.5$ and other couplings defined in the Appendix   as $\alpha'_2 \lesssim 30$, and $y'_1,\, z' \lesssim 1$. The vevs $\kappa_1$ is taken as a free parameters in the range of [53, 220] GeV (the two extreme values are determined by requiring that the Yukawa coupling in the quark mass matrices are not too large); we choose $\kappa_2 = \kappa_1/5$ (this ratio respects our numerical fit of the $4\times4$ up-type mass matrix with CP violation) and $v_\rho = \sqrt{v^2 - \kappa_1^2 - \kappa_2^2}$. We select the sets of parameters which predict the lightest Higgs mass $m_h \in [123,\,127]$ GeV. For the pseudoscalar mass plot, we set explicitly $M' = v$, and $0 < \alpha_{2,\,6} \lesssim 3$.}.
In the fit we found that $\alpha'_2$ is required to be large $\sim10$, as we argued above. It should be stressed, however, that a large $\alpha'_2$ does not necessarily mean the invalidity of perturbation theory, since it is the the linear summation of five independent scalar self couplings, cf. Eq.~(\ref{parameter-prime}). Furthermore, to get a 125 GeV Higgs, $\kappa_1$ is found to be large, $\sim(130,\,210)$ GeV, which means that generally both Re$\phi_1^0$ and Re$\rho_0$ contribute substantially to the SM Higgs.

As an explicit example, when we set the vevs
\begin{eqnarray}
&& \kappa_1 = 183 \, {\rm GeV} \,, \nonumber\\
&& \kappa_2 = 36.6 \, {\rm GeV} \,, \nonumber\\
&& v_\rho = 160 \, {\rm GeV} \,, \nonumber\\
&& v_R = 3305 \, {\rm GeV} \,,
\end{eqnarray}
and the quartic couplings (all defined in Appendix B)
\begin{eqnarray}
&& \alpha'_2 = 10.1 \,, \nonumber\\
&& \gamma' = 0.5 \,, \nonumber \\
&& y'_1  = 0.588 \,, \nonumber\\
&& z' = 0.064 \,,
\end{eqnarray}
we can get the SM Higgs (the lightest one) with mass 125 GeV, with help of the rotation matrix connecting the flavor states and the mass eigenstates (the superscript $S$ stands for ``scalar'')
\begin{eqnarray}
\left(\begin{matrix} h \\ H_1 \\ H_2 \\ H_R \end{matrix}\right)
= U^{S}
\left(\begin{matrix}
{\rm Re}\phi^0_1 \\ {\rm Re}\phi^0_2 \\
{\rm Re}\rho^0 \\ {\rm Re}\delta^0_R
\end{matrix}\right)
\end{eqnarray}
with values
\begin{eqnarray}
U^{S} = \left( \begin{array}{cccc}
 0.762 & 0.086 & 0.640 & -0.029 \\
 0.624 & -0.354 & -0.695 & -0.007 \\
 0.166 & 0.930 & -0.325 & -0.041 \\
 0.0339 & 0.0386 & 0.00065 & 0.998 \\
\end{array} \right) \,.
\end{eqnarray}
(Note that this choice of the $\kappa_{1,2}$ gives the realistic CP violating CKM mixings cf. line after Eq. (\ref{No.25}).) In this case, the masses of the three heavier states are, respectively,
\begin{eqnarray}
&& m_{H_1} = 812  \, {\rm GeV} \,, \nonumber \\
&& m_{H_2} = 1335 \, {\rm GeV} \,, \nonumber \\
&& m_{H_R} = 3309 \, {\rm GeV} \,.
\end{eqnarray}

\section{FCNH effects Constraint on the second Higgs boson mass}
To investigate the FCNH constraints, we look at the couplings of the neutral Higgs fields in the Yukawa coupling given in Eq. (5).
For the down-type quarks, all the couplings to the scalars are proportional to the diagonal $\mathcal{M}_d$ and therefore they do not introduce no flavor changing neutral couplings. However, for the up-type quarks, the situation is completely different. The neutral Higgs couplings for the up sector can be written as:
\begin{eqnarray}
-\mathcal{L}^0_Y 
   &=& \bar{u}_L (h_q \phi^0_1) u_R
   + \bar{u}_L (h_f \rho^{0 \ast}) t'_R
   + \bar{t'}_L (h_g \delta^0_R) u_R
   + M \bar{t'}_L t'_R \,.
\end{eqnarray}
Using the fact that the vevs of these fields contribute to up quark mass matrix $\mathcal{M}_u$, we can write $\mathcal{L}^0_Y$ as:
\begin{eqnarray}
-\mathcal{L}^0_Y &=& \bar{\mathcal{U}}_L \mathcal{M}_u \mathcal{U}_R
  + (\sqrt2 G_{\rm F})^{1/2}
  \bar{\mathcal{U}}_L \mathcal{M}'_u h \mathcal{U}_R
  + (\sqrt2 G_{\rm F})^{1/2}
  \bar{\mathcal{U}}_L \mathcal{M}^{\prime\prime}_u H_1 \mathcal{U}_R
  + \cdots \,,
\end{eqnarray}
where
\begin{eqnarray}
\mathcal{U} = \left( \begin{matrix} u_a \\ t' \end{matrix} \right)
\end{eqnarray}
and
\begin{eqnarray}
&& \mathcal{M}'_u =
\left( \begin{matrix} c'_0 h & c'_1 f \\ c'_2 g & 0 \end{matrix} \right) \,, \\
&& \mathcal{M}''_u =
\left( \begin{matrix} c''_0 h & c''_1 f \\ c''_2 g & 0 \end{matrix} \right) \,.
\end{eqnarray}
In the numerical example given above,
\begin{equation}
\begin{array}{ll}
  c'_0  = \frac{v}{\kappa_1} U^S_{11} = 1.025 \,,
& c''_0 = \frac{v}{\kappa_1} U^S_{21} = 1.58 \,,  \\
  c'_1  = \frac{v}{v_\rho} U^S_{13} =  0.984 \,,
& c''_1 = \frac{v}{v_\rho} U^S_{23} =  -0.620 \,,\\
  c'_2  = \frac{v}{v_R} U^S_{14} = - 0.0022 \,,
& c''_2 = \frac{v}{v_R} U^S_{24} = - 0.00093 \,.
\end{array}
\end{equation}

When we transform the up-type quarks into their mass eigenstates, the coupling of SM higgs $h$ couples to the quarks via the matrix
\begin{eqnarray}
V_{\rm L} \mathcal{M}'_u V^\dagger_{\rm R} \simeq
\left(
\begin{array}{cccc}
 0.0301 & 0.000074 & -0.00098 & -0.0736 \\
 -0.000083 & 0.543 & -0.00323 & -0.372 \\
 -0.176 & 0.497 & 148. & -49.0 \\
 -0.0513 & 0.144 & -4.36 & -5.20 \\
\end{array}
\right) \,,
\end{eqnarray}
where $V_L$ and $V_R$ are, respectively, the left- and right-handed quark mixing matrices. Here follow two comments:
\begin{itemize}
  \item
  It is transparent from the left panel of Fig.~\ref{ytop+ycu} that the coupling to top quark (and the up and charm quarks, given potential corrections to the up quark Yukawa coupling) is mostly about the same as in the SM (note that when evaluated at TeV scale, $m_t\simeq150$ GeV), and the coupling to the fourth heavy quark is tiny and its contribution to Higgs production at LHC can be safely neglected, with a contribution of $\sim10^{-3}$ to $10^{-4}$.
  \begin{figure}[t]
    \begin{center}
    \includegraphics[height=4.5cm]{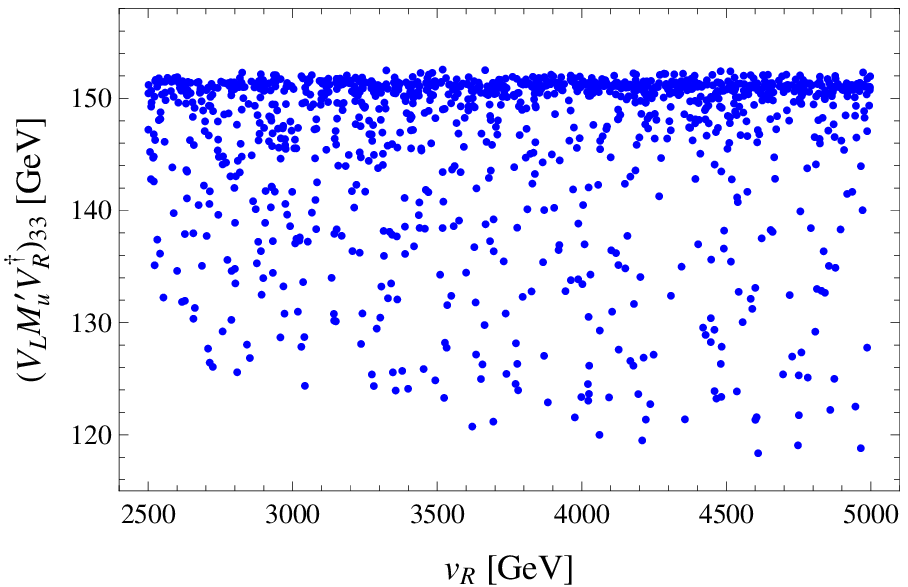} \hspace{.2cm}
    \includegraphics[height=4.5cm]{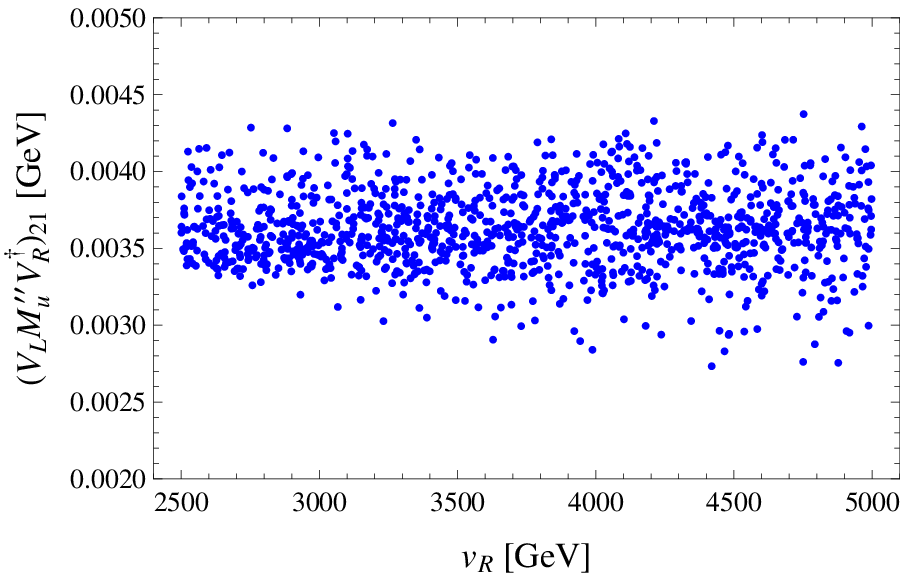}
    \vspace{-.2cm}
    \caption{Distributions of $(V_L \mathcal{M}'_u V_R^\dagger)_{33}$ and $(V_L \mathcal{M}''_u V_R^\dagger)_{21}$ as functions of $v_R$, in unit of GeV.}
    \vspace{-.7cm}
    \label{ytop+ycu}
    \end{center}
  \end{figure}
  \item
  It seems that the flavor-changing coupling of the SM Higgs is very small, the largest one of which being the element involving the flavor changing Higgs coupling to the top and charm quark (we do not consider the elements relevant to the heavy fourth quark). The element $(V_L \mathcal{M}'_u V_R^\dagger)_{32}$ imply a Yukawa coupling $y_{tch}$ of order $10^{-4} - 10^{-2}$, as shown in Fig.~\ref{tch}. There has been much discussion on constraints of such flavor-changing coupling from top and Higgs experimental data~\cite{fcnc-general,fcnc-craig,fcnc-chen}. The ATLAS collaboration recently got an upper limit of per cent level on the non-standard top decay $t \rightarrow hc$ channel~\cite{fcnc-atlas}, which imply an upper limit of order 0.1 on the Yukawa coupling $y_{tch}$. The authors of~\cite{fcnc-gupta} assert that the processes $pp \rightarrow t\bar{j}h,\, \bar{t}jh$ at LHC can produce a more stringent constraint, of the order of $10^{-3}$. As far as our model goes, it is evident from Fig.~\ref{tch} that there exists a large region of parameter space which respects the experimental constraints on the flavor changing top couplings.
  \begin{figure}[t]
    \begin{center}
    \includegraphics[width=7cm]{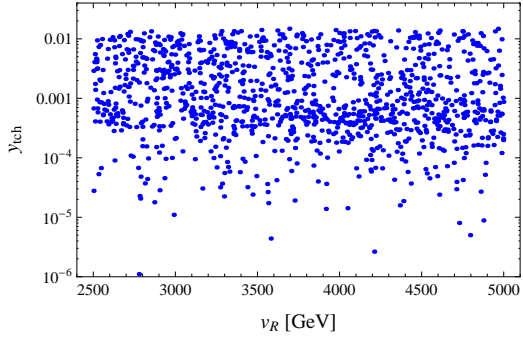} 
    \vspace{-.3cm}
    \caption{Flavor changing Yukawa coupling $y_{tch}$ as function of $v_R$.}
    \vspace{-.4cm}
    \label{tch}
    \end{center}
  \end{figure}
\end{itemize}

In the example, the coupling to the second Higgs $H_1$ is proportional to
\begin{eqnarray}
V_{\rm L} \mathcal{M}''_u V^\dagger_{\rm R} \simeq
\left(
\begin{array}{cccc}
 0.0235 & 0.003 & -0.0408 & -0.0475 \\
 -0.0036 & 0.455 & -0.125 & -0.265 \\
 -8.18 & 23.0 & -159 & 49.4 \\
 0.123 & -0.345 & 2.19 & -2.24 \\
\end{array}
\right) \,.
\end{eqnarray}
the term relevant to $D$ meson mixing is very small
\begin{eqnarray}
(V_L \mathcal{M}''_u V_R^\dagger)_{21} \simeq -3.6\,{\rm MeV} \,.
\end{eqnarray}
Then the effective Lagrangian reads,
\begin{eqnarray}
\mathcal{L}_{\rm eff} \sim \frac{G_{\rm F}}{m^2_{H_1}}
(V_L \mathcal{M}''_u V_R^\dagger)_{21}^2
\Big[ (c\bar{u})^2 -(c\gamma_5 \bar{u})^2 \Big] \,,
\end{eqnarray}
which produces a extremely small contribution to $D^0 - \bar{D}^0$ mixing,
\begin{eqnarray}
\Delta {m}_{D}
&\sim& \langle \bar{D} | \mathcal{L}_{\rm eff} | D \rangle \nonumber \\
&\sim& \frac{G_{\rm F}}{m^2_{H_1}} (V_L \mathcal{M}''_u V_R^\dagger)_{21}^2 m_D F_D^2
\left( \frac{m_D}{m_c} \right)^2 \nonumber \\
&\sim& 2.5\times10^{-17} \left( \frac{\rm TeV}{m_{H_1}} \right)^{2} \, {\rm GeV} \,.
\end{eqnarray}
The experimental value of the mass difference is of order $10^{-14}\,{\rm GeV}$; this therefore does not severely constrain the  $H_1$ mass. The right panel of Fig.~\ref{ytop+ycu} shows that the flavor-changing contribution to $D^0-\bar{D}^0$ mixing is stable against the variation of the parameters in the potential and variation of $v_R$. One might na\"ively expect that $(V_L \mathcal{M}''_u V_R^\dagger)_{21}$ should be dominated by the term $\sim m_c (V_R^\dagger)_{21}$, however, numerical data shows that this terms is always cancelled by other contributions, e.g. the terms proportional to $(\mathcal{M}''_u)_{33}$ or $(\mathcal{M}''_u)_{34}$, leaving a very tiny contribution. As stated in the introduction, this is guaranteed by the fact that the the SM quark mixing is produced by tiny mixing with the vector-like heavy fermion.

\section{LHC prospects for  $H_1$}

\begin{figure}[t]
  \begin{center}
  \includegraphics[width=7.5cm]{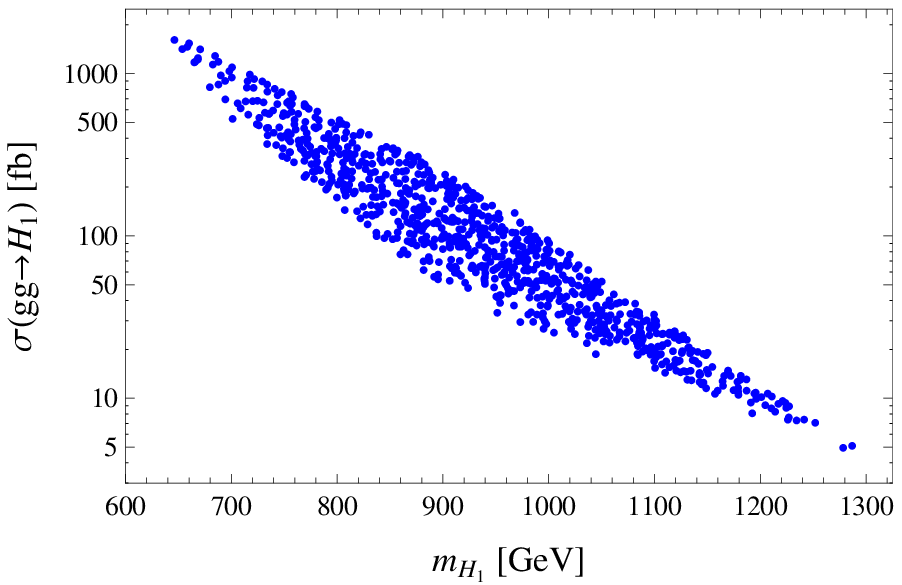}
  \includegraphics[width=7.5cm]{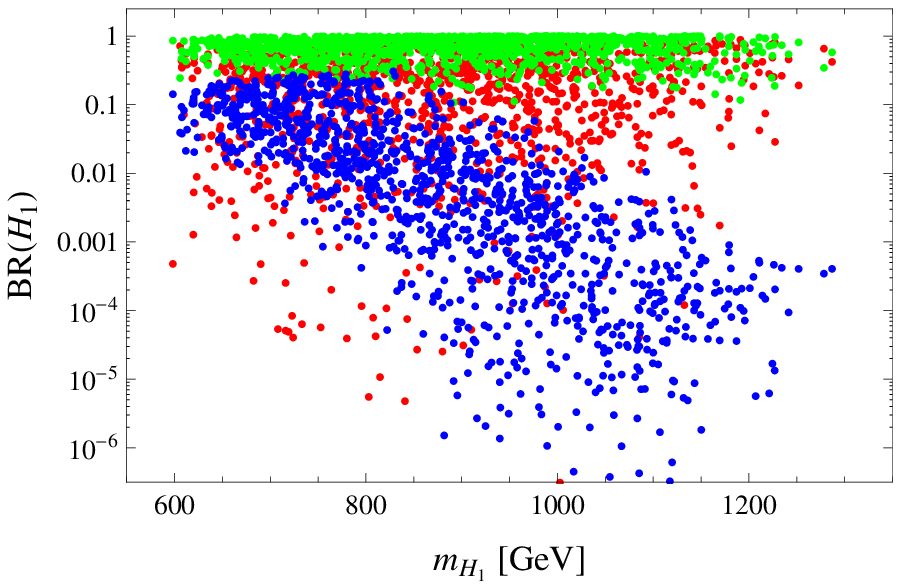}
  \vspace{-.2cm}
  \caption{Left panel: $H_1$ production cross section $\sigma (gg \rightarrow H_1)$ at LHC with center-of-mass energy of 14 TeV, as function of $H_1$ mass. Right panel: Branching ratios of $H_1$ decay, as function of its mass. The decay channels $H_1 \rightarrow hh,\, t\bar{t},\, WW$ are shown, respectively, as red, green and blue spots.}
  \vspace{-.7cm}
  \label{production+decay}
  \end{center}
\end{figure}

In our model, $H_1$ behave somewhat like a heavy copy of the SM Higgs: it has hierarchical couplings to the SM fermions, and couples also to the gauge bosons. Numerical analysis reveals that its coupling to the top quark is generally of order one; therefore, similarly to the SM Higgs, the top-loop induced gluon fusion process would be the dominant production channel for $H_1$ at the LHC. It is very straightforward to obtain the leading order cross section for the production process $gg \rightarrow H_1$: we need only to rescale that for the SM Higgs (extrapolated to $\sim$ TeV mass) by the Yukawa coupling ratio $(V_L \mathcal{M}''_u V_R^\dagger)_{33}^2 / (V_L \mathcal{M}'_u V_R^\dagger)_{33}^2$. The left panel of Fig.~\ref{production+decay} depicts the sizable production cross section and its dependence on the scalar mass. It is evident in the figure that, for a TeV $H_1$, the cross section can be as large as 100 fb at $\sqrt{s}=  14$ TeV, and we could expect thousands of $H_1$ events produced at upgraded LHC. For a lighter $H_1$, the cross section could be even larger.

We now sketch how the second Higgs decays after its production. Once the scalars obtain their non-zero vevs, there appear cubic couplings among the Higgs states, which in the flavor basis are given by,
\begin{eqnarray}
&& \Big[ \alpha'_1 \kappa_1 ({\rm Re}\phi_1^0)^3
+  \alpha'_1 \kappa_2 ({\rm Re}\phi_2^0)^3
+  \beta' v_\rho ({\rm Re}\rho^0)^3
+  \gamma' v_R ({\rm Re}\delta_R^0)^3 \Big] \nonumber \\
&& +\frac12 \Big[
     x'_1 v_\rho ({\rm Re}\phi_1^0)^2 ({\rm Re}\rho^0)
    +x'_1 \kappa_1 ({\rm Re}\phi_1^0) ({\rm Re}\rho^0)^2
    +x'_2 v_\rho ({\rm Re}\phi_2^0)^2 ({\rm Re}\rho^0)
    +x'_2 \kappa_2 ({\rm Re}\phi_2^0) ({\rm Re}\rho^0)^2
    \Big] \nonumber \\
&& +\frac12 \Big[
     y'_1 v_R ({\rm Re}\phi_1^0)^2 ({\rm Re}\delta_R^0)
    +y'_1 \kappa_1 ({\rm Re}\phi_1^0) ({\rm Re}\delta_R^0)^2
    +y'_2 v_R ({\rm Re}\phi_2^0)^2 ({\rm Re}\delta_R^0)
    +y'_2 \kappa_2 ({\rm Re}\phi_2^0) ({\rm Re}\delta_R^0)^2
    \Big] \nonumber \\
&& +\frac12 \Big[
     z' v_R ({\rm Re}\rho^0)^2 ({\rm Re}\delta_R^0)
    +z' v_\rho ({\rm Re}\rho^0) ({\rm Re}\delta_R^0)^2
    \Big] \nonumber \\
&& -\frac{1}{\sqrt2} \Big[
     M' ({\rm Re}\phi_1^0) ({\rm Re}\rho^0) ({\rm Re}\delta_R^0)
    +M' ({\rm Re}\phi_2^0) ({\rm Re}\rho^0) ({\rm Re}\delta_R^0)
    \Big].
\end{eqnarray}
After rotating these states to the physical basis, we can easily get the dimensionful cubic coupling $m_{H_1 hh} H_1 hh$ from the potential. To calculate the decay width $\Gamma(H_1 \rightarrow hh)$, we assume all the relevant original couplings in the potential (those which are not constrained by the observation that $m_h = 125$ GeV) lie in the perturbative range\footnote{If these parameters have much larger values, e.g. $\lesssim 4\pi$, then in a large portion of parameter space, the decay channel $H_1 \rightarrow hh$ dominates over others. However, here we consider the preferable smaller values of couplings, $\sim(0,\,3)$ in giving the branching ratios. 
In numerical calculations we take values in the exact range [0, $\sqrt{4\pi}$].} $\sim(0,\,3)$. This leads to the decay width as:
\begin{eqnarray}
\Gamma (H_1 \rightarrow hh) =
\frac{1}{8\pi} \frac{m_{H_1 hh}^2}{m_{H_1}}
\left( 1 - \frac{4m_h^2}{m_{H_1}^2} \right)^{1/2} \,.
\end{eqnarray}
Turning to the fermion decay channels, the top quark mode dominates over others. As for the coupling of $H_1$ to top quark, it is easy to see that the Yukawa coupling $y_{H_1 t\bar{t}} = \sqrt2 (V_L M''_u V_R^\dagger)_{33} / v$, which produces the fermionic decay width
\begin{eqnarray}
\Gamma (H_1 \rightarrow t\bar{t}) =
\frac{3}{16\pi} \cdot \left| y_{H_1 t\bar{t}} \right|^2 {m_{H_1}}
\left( 1 - \frac{4m_t^2}{m_{H_1}^2} \right)^{3/2} \,.
\end{eqnarray}
Besides the scalars and fermions $H_1$ can also decay into SM gauge bosons, $WW,\, ZZ$, but those widths are generally suppressed by the smaller gauge coupling. For instance, for $H_1 \rightarrow WW$, we define $m_{H_1 WW} = U_{21} \kappa_1 + U_{22} \kappa_2 + U_{23} v_\rho$ and $f_W = \frac12 g^2 m_{H_1 WW}$,
then the width reads
\begin{eqnarray}
\Gamma (H_1 \rightarrow WW) =
\frac{1}{8\pi} \frac{f_W^2}{m_{H_1}}
\left[ 1 + \frac12 \left( 1 - \frac{m_{H_1}^2}{2m^2_W} \right)^2 \right]
\left( 1 - \frac{4m_W^2}{m_{H_1}^2} \right)^{1/2} \,.
\end{eqnarray}
The $ZZ$ channel is expected to have similar width. As an explicit numerical example, the branching ratios of these different decay channels are shown in the right panel of Fig.~\ref{production+decay}. Obviously, in almost all the parameter space, the top quark channel dominates as mentioned before , enhanced by the order one Yukawa coupling.

\section{Lepton sector}
Let us briefly comment on the lepton sector of the model. It is clear from Eq. (5) that $M_\ell = r M_D$, where $M_{\ell, D}$ are the charged lepton and neutrino Dirac masses. However, due to this fact and the fact that $v_R \sim 3$ TeV, simple generic type I seesaw cannot reproduce the observed neutrino spectrum and mixings. We have tried possible new textures for $y_R$ matrix but have not found any that will help us to get right neutrino mass pattern. It seems that the simples way to accommodate small neutrino masses in our scheme is to add three gauge singlet neutrinos, a right handed doublet Higgs field and invoke inverse seesaw mechanism~\cite{inverse}. According to the inverse seesaw paradigm, the right handed neutrino of the left-right model is a quasi-Dirac neutrino and its characteristic signature is a trilepton final state rather than  the $\ell^\pm\ell^\pm jj$ mode. This implies that the search strategy for the $W_R$ at LHC must change in case a TeV scale heavy Higgs  is found.  This situation can also have interesting implicatios for neutrinoless double beta decay~\cite{parida1}.

\section{Other phenomenological and theoretical comments}

(i) In addition to the three near TeV LHC accessible neutral Higgs fields, two new fermions are the top partner $t'$ and the $5/3$ charged quarks with new phenomenology. The presence of the $5/3$ charged quark also alters the phenomenology of the doubly charged Higgs bosons~\cite{france}. From our analysis, it appears that the CKM fit requires $m_{t'}\sim 3$ TeV. Since we expect $m_{t'}\simeq m_{Q_{5/3}}$ due to the vector-like nature of the extra fermion doublet, both the $t'$ and $Q_{5/3}$ could be accessible at the LHC. As was noted in~\cite{france}, unlike the decay properties of $Q_{5/3}$ discussed in the literature so far~\cite{q53}, where it is assumed that $Q_{5/3}\to t+W^+$, in our scheme (as in the scheme in ~\cite{france}), the primary decay mode of $Q_{5/3}$ could be $Q_{5/3}\to \Delta^{++}+d$ and/ or $Q_{5/3}\to \rho^{++}+d$  leading to different LHC signal.
\vspace{0.2cm}

(ii) The model has also new charged scalar states beyond that of the minimal LRSM. For instance there are the new doubly charged fields fields $\rho^{++}$, whose mass is expected to be in the TeV range.  Its likely decay mode is $\rho^{++}\to \Delta^+\phi^+_2$ with $\Delta^+$ subsequently decaying to leptonic final states. There are also new singly charged states as partners of $H_{1,2}$ with associated rich phenomenology as are the pseudo-scalar partners of $H_{1,2}$. We do not pursue this any further in this paper.
\vspace{.2cm}

(iii) An interesting point about our model is that even though the SM Higgs coupling to $b\bar{b}$ arises in an indirect way compared to SM (via coupling to $\phi^0_2$ and fraction of SM Higgs in $\phi^0_2$), its magnitude remains almost same as in SM.
\vspace{.2cm}

(iv) It is also worth noting that one could envision adding more than one vector-like quark multiplets (instead of one that we have added). Such models would lead to changes in the details of the model e.g. Higgs decays, masses etc. The model discussed here should be considered as an existence of proof of models with lower Higgs mass in left-right symmetric models rather than as a definite final model.
\vspace{.2cm}

(v) Finally, a theoretical comment on the model: the new fermion and scalar multiplets we have chosen to include can emerge from an SO(10) grand unified theory. For instance, the fermion $Q'$ and Higgs fields $\rho$ added to minimal LRSM could also have grand unified origin since the $Q'$ is part of the {\bf 560}+${\bf 560^*}$ representation whereas the $\rho$-Higgs field is a sub-multiplet of {\bf 210} representation. The new Yukawa couplings involving $Q'$ and $\rho$ in Eq. (5), $y_f$ and $y_g$ couplings can arise from SO(10) invariant couplings ${\bf  16\cdot 210 \cdot 560^*}$ and ${\bf  16\cdot 126^*\cdot 560}$ in a possible GUT version of the theory. We have also checked that if there is a complex $Y=0$ $SU(2)_L$ triplet in the theory at the TeV scale, three couplings of SM almost unify at the one loop level via the chain $G_{SM}\to G_{LRSM}\to SO(10)$ with $G_{LRSM}$ at the TeV scale. Given that we have not included threshold effects as well as two loop contributions, this unification is likely to be better. Such triplets could be the ones present in {\bf 45} or {\bf 54} Higgs fields that are sued to break the SO(10) symmetry.  We assume that all the other fields in the fermion multiplet {\bf 560} except the $Q'$ must be at the GUT scale. We do not pursue these and other  detailed aspects of grand unification possibility any further since it is beyond the scope of this paper.

\section{Summary} 
In this paper, we have explored the question of the second Higgs mass in the left-right symmetric models. The well known fact that in the minimal version of this model, the second neutral Higgs mass is more than 12-15 TeV implies that discovery of a neutral Higgs with a few TeV mass ($\ll$ 10 TeV) would rule out the minimal LRMS. The question would then be: should the search for $W_R$ at LHC continue in this case?  In other words, are there versions of the left-right model that keep the $W_R$ mass in the 5-6 TeV range while at the same time having a Higgs mass of a few TeV or less without conflicting with meson-anti-meson constraints ($K^0-\bar{K}^0$, $B^0-\bar{B}^0$ mixing etc)? We have provided an existence proof of such models in this paper  and what their basic features should be. It appears from our example that such models should have heavier vector like quarks and/or extra Higgs fields to generate desired CKM mixings. An interesting consequence of these models also appears to be that the search mode for $W_R$ at the LHC should change to $\ell^\pm\ell^\mp\ell^\pm$ + missing E rather than $\ell^\pm\ell^\pm jj$ with no missing E, which is currently being pursued~\cite{gin}. In the particular example we provide, the heavier neutral Higgs masses are all below the $W_R$ mass, although their precise mass values could shift in this range depending on model details. Needless to say that search for a TeV scale Higgs is therefore crucial for understanding the left-right symmetric extensions of the standard model and how neutrino masses arise in such models.

\section*{Acknowledgement}
We like to thank Goran Senjanovi\'c and Yue Zhang for discussions. The work of R. N. M. is supported by the National Science Foundation grant No. PHY-1315155. This work of Y. Z. is supported in part by the National Natural Science Foundation of China (NSFC) under Grant No. 11105004.

\vspace{0.3in}
\appendix
\section{CKM fit without CP violation}
In this appendix, as a toy test of our model, we consider the fit of CKM matrix without any CP violation. Counting the numbers of parameters in Eq.~(\ref{diagonalization}): on the LHS, we have eight parameters $r$, $f_{1,\,2,\,3}$, $g_{1,\,2,\,3}$ and $M$, whereas on the RHS, we have ten to be fit, the four quark masses $m_{u,\,c,\,t}$, $m'_t$ and the six independent mixing angles $\theta_{12,\,23,\,13,14,24,34}$. Of the ten parameters on the RHS, six are known and the other four ($m'_t$ and $\theta_{14,\,24,\,34}$) involve new physics are yet undetermined. To find reasonable solutions for our model, we need to fix two of the four unknown parameters on the RHS. Explicitly, we choose\footnote{We find that if we set $m'_t$ at lower values, say 1 TeV, then, due to mixing of $t'$ with the SM fermions, the values of $M$ would be much smaller than the TeV scale (e.g., for $m'_t$ = 1 TeV, $M\sim$ 10 GeV), which seems less appealing. Thus we choose a somewhat larger $t'$ quark mass.}
\begin{eqnarray}
\label{choice}
&& m'_t = 3\,{\rm TeV} \,, \nonumber \\
&& \theta_{34} = 0.016 \,.
\end{eqnarray}
The choice of the value for $\theta_{34}$ means that it lies on the $2\sigma$ boundary of the SM constraint.

A typical solution for the parameters that fits all quark masses and mixings  is given below (all masses in unit of GeV):
\begin{equation}
\begin{array}{llll}
\vspace{.1cm}
r = 10.1 \,, &
f_1 = -0.916\,, &
g_1 = -465 \,, &
M = 668 \,. \\
\vspace{.1cm}
&
f_2 = 6.26 \,, &
g_2 = -2260 \,, & \\
&
f_3 = -157 \,, &
g_3 = -1797 \,, &
\end{array}
\end{equation}
 The resulting $4\times 4$ quark mixing matrix reads
\begin{eqnarray}
\left(
\begin{array}{cccc}
 0.974 & 0.225 & 0.0035 & 0.000047 \\
 -0.225 & 0.973 & 0.041 & 0.00032 \\
 0.0058 & -0.0409 & 0.999 & 0.016 \\
 -0.000069 & 0.00034 & -0.0165 & 0.999 \\
\end{array}
\right) \,,
\end{eqnarray}
which fits very well the $3\times3$ CKM matrix and it is transparent that the mixings between the SM up-type quark with the 4th one are very small.
\begin{eqnarray}
&& s_{14} = 0.000047 \,, \nonumber \\
&& s_{24} = 0.00032 \,, \nonumber \\
&& s_{34} = 0.016 \,,
\end{eqnarray}
and it is reasonable that the mixing with the top quark is comparatively much larger than with the first two generations.

\section{Potential and scalar mass-squared matrices}
\subsection{Potential and minimization conditions}
In this appendix, we discuss the Higgs potential of our model, its minimization to get the mass eigenstates and eigenvalues for the CP even and CP odd real scalar fields of the model. The full Higgs potential of the model invariant under  the discrete and gauge symmetries is:
\begin{eqnarray}
\label{potential}
V &=& -\mu_1^2 {\rm Tr} (\phi^{\dag} \phi)
      +M_\rho^2 {\rm Tr} (\rho^{\dag} \rho)
      -\mu_3^2 {\rm Tr} (\Delta_R \Delta_R^{\dag})
    \nonumber  \\
&& +\alpha_1 \left[ {\rm Tr} (\phi^{\dag} \phi) \right]^2
   +\alpha_2 \left\{ \left[ {\rm Tr} (\tilde{\phi} \phi^{\dag}) \right]^2 + \left[ {\rm Tr} (\tilde{\phi}^{\dag} \phi) \right]^2 \right\}
   +\alpha_3 {\rm Tr} (\tilde{\phi} \phi^{\dag}) {\rm Tr} (\tilde{\phi}^{\dag} \phi)
   \nonumber \\
&& +\alpha_4 {\rm Tr} (\phi^\dagger \phi \phi^\dagger \phi)
   +\alpha_5 {\rm Tr} (\phi^\dagger \phi \tilde\phi^\dagger \tilde\phi)
   +\alpha_6 \left[ {\rm Tr} (\phi^\dagger \tilde\phi \phi^\dagger \tilde\phi)
   +{\rm Tr} (\phi \tilde\phi^\dagger \phi \tilde\phi^\dagger) \right]
   \nonumber \\
&& +\beta_1 \left[ {\rm Tr} (\rho^{\dag} \rho) \right]^2
   +\beta_2 {\rm Tr} (\tilde{\rho} \rho^{\dag}) {\rm Tr} (\tilde{\rho}^{\dag} \rho)
   +\beta_3 {\rm Tr} (\rho^{\dag} \rho \rho^{\dag} \rho)
   +\beta_4 {\rm Tr} (\rho^{\dag} \rho \tilde\rho^{\dag} \tilde\rho)
   \nonumber \\
&& +\gamma_1 \left[ {\rm Tr} (\Delta_R \Delta_R^{\dag}) \right]^2
   +\gamma_2 {\rm Tr} (\Delta_R \Delta_R) {\rm Tr} (\Delta_R^\dagger \Delta_R^\dagger) \nonumber \\
&& +\gamma_3 {\rm Tr} (\Delta_R \Delta_R^{\dag} \Delta_R \Delta_R^{\dag})
   +\gamma_4 {\rm Tr} (\Delta_R \Delta_R \Delta_R^{\dag} \Delta_R^{\dag})
   \nonumber \\
&& +x_1 {\rm Tr} (\phi^\dagger \phi) {\rm Tr} (\rho^\dagger \rho)
   +x_2 {\rm Tr} (\phi^\dagger \phi \rho^\dagger \rho)
   +x_3 {\rm Tr} (\phi \phi^\dagger \rho \rho^\dagger) \nonumber \\
&& +x_4 {\rm Tr} (\tilde\phi \tilde\phi^\dagger \rho \rho^\dagger)
   +x_5 {\rm Tr} (\tilde\phi^\dagger \tilde\phi \rho^\dagger \rho)
    \nonumber \\
&& +x_6 \left[ {\rm Tr} (\tilde\phi \tilde\rho^\dagger \phi \rho^\dagger) + {\rm Tr} (\tilde\rho \tilde\phi^\dagger \rho \phi^\dagger) \right]
   +x_7 \left[ {\rm Tr} (\phi \tilde\rho^\dagger \tilde\phi \rho^\dagger) + {\rm Tr} (\rho \tilde\phi^\dagger \tilde\rho \phi^\dagger) \right]
   \nonumber \\
&& +y_1 {\rm Tr} (\phi^\dagger \phi) {\rm Tr} (\Delta_R \Delta_R^\dagger)
   +y_2 {\rm Tr} (\phi^\dagger \phi \Delta_R^\dagger \Delta_R)
   +y_3 {\rm Tr} (\phi^\dagger \phi \Delta_R \Delta_R^\dagger)
   \nonumber \\
&& +z_1 {\rm Tr} (\rho^\dagger \rho) {\rm Tr} (\Delta_R \Delta_R^\dagger)
   +z_2 {\rm Tr} (\rho^\dagger \rho \Delta_R^\dagger \Delta_R)
   +z_3 {\rm Tr} (\rho^\dagger \rho \Delta_R \Delta_R^\dagger) \,.
\end{eqnarray}
To get non-vanishing vev of the extra $\rho$ scalar, we introduce the terms soft breaking the discrete symmetry,
\begin{eqnarray}
\label{soft}
M' \left[
 {\rm Tr} (\tilde\rho \Delta_R \phi^\dagger)
+{\rm Tr} (\tilde\rho \Delta_R \tilde\phi^\dagger)
+{\rm h.c.} \right] \,.
\end{eqnarray}
For simplicity, we assume all the couplings in the Lagrangian are real parameters.

The four minimization conditions are:
\begin{eqnarray}
 \frac{\partial}{\partial \kappa_1} V
=\frac{\partial}{\partial \kappa_2} V
=\frac{\partial}{\partial v_\rho} V
=\frac{\partial}{\partial v_R} V = 0.
\end{eqnarray}
They lead to the following relations among the vevs and the coefficient in the potential,
\begin{eqnarray}
&& \frac{\mu^2_1}{v^2_R} =
  \frac{y'_1}{2}
+ \alpha'_1 \frac{\kappa^2_1}{v^2_R}
+ \alpha'_2 \frac{\kappa^2_2}{v^2_R}
+ \frac{x'_1}{2} \frac{v^2_\rho}{v^2_R}
- \frac{1}{\sqrt2} \frac{M'}{\kappa_1} \frac{v_\rho}{v_R} \,,
\nonumber \\
&& \frac{\mu^2_1}{v^2_R} =
  \frac{y'_2}{2}
+ \alpha'_2 \frac{\kappa^2_1}{v^2_R}
+ \alpha'_1 \frac{\kappa^2_2}{v^2_R}
+ \frac{x'_2}{2} \frac{v^2_\rho}{v^2_R}
- \frac{1}{\sqrt2} \frac{M'}{\kappa_2} \frac{v_\rho}{v_R} \,,
\nonumber \\
&& \frac{1}{\sqrt2} \frac{\kappa_1 + \kappa_2}{v_R} \frac{M'}{v_\rho}
   -\frac{M^2_\rho}{v^2_R} =
  \frac{z'}{2}
+ \frac{x'_1}{2} \frac{\kappa^2_1}{v^2_R}
+ \frac{x'_2}{2} \frac{\kappa^2_2}{v^2_R}
+ \beta' \frac{v^2_\rho}{v^2_R} \,,
\nonumber \\
&& \frac{\mu^2_3}{v^2_R} =
  \gamma'
+ \frac{y'_1}{2} \frac{\kappa^2_1}{v^2_R}
+ \frac{y'_2}{2} \frac{\kappa^2_2}{v^2_R}
+ \frac{z'}{2} \frac{v^2_\rho}{v^2_R}
 -\frac{1}{\sqrt2} \frac{M'}{v_R} \frac{\kappa_1 + \kappa_2}{v_R} \frac{v_\rho}{v_R} \,,
\end{eqnarray}
where
\begin{equation}
\label{parameter-prime}
\begin{array}{ll}
\vspace{.1cm}
\alpha'_1 = \alpha_1 +\alpha_4 \,, &
x'_1 = x_1 +x_2 +x_4 -2x_6 \,,  \\
\vspace{.1cm}
\alpha'_2 = \alpha_1 +4\alpha_2 +2\alpha_3 +\alpha_5 +2\alpha_6 \,, &
x'_2 = x_1 +x_3 +x_5 -2x_7 \,,  \\
\vspace{.1cm}
\beta' = \beta_1 +\beta_3 \,, &
y'_1 = y_1 +y_2 \,, \\
\vspace{.1cm}
\gamma' = \gamma_1 +\gamma_3 \,, &
y'_2 = y_1 +y_3 \,, \\
&
z'   = z_1 +z_2 \,.
\end{array}
\end{equation}
Assuming the two mass parameters $M_\rho,\,M' \sim v_{EW}$, we find
at leading order of $\kappa_1,\,\kappa_2,\,v_\rho \ll v_R$ that these these relations are greatly simplified,
\begin{eqnarray}
\label{parameter-LO}
&& \frac{\mu^2_1}{v^2_R} \cong
  \frac{y_1 +y_2}{2} \,, \nonumber \\
&& \frac{\mu^2_1}{v^2_R} \cong
  \frac{y_1 +y_3}{2} \,, \nonumber \\
&& \frac{1}{\sqrt2} \frac{\kappa_1 + \kappa_2}{v_R} \frac{M'}{v_\rho} \cong
  \frac{z_1 +z_2}{2} \,, \nonumber \\
&& \frac{\mu^2_3}{v^2_R} \cong
  \gamma_1 +\gamma_3 \,.
\end{eqnarray}
The first two equations imply (at the leading order of $\frac{\kappa_{1,\,2},\,v_\rho}{v_R}$) $y_2 \simeq y_3$. There are two equations (the first and fourth ones) for the vev $v_R$, indicating {fine-tuning} of some of the coefficients, as in the case of manifest minimal LRSM~\cite{Zhang:2007da}.

\subsection{Neutral Higgs boson mass-squared matrices}
Using the above conditions, we find that
in the basis of $\{ {\rm Re}\phi^0_1,\, {\rm Re}\phi^0_2,\, {\rm Re}\rho^0,\, {\rm Re}\delta^0_R \}$, the neutral Higgs mass-square matrix elements read~\cite{Deshpande:1990ip}
\begin{eqnarray}
&& M^{2\rm Re}_{11} =
-\mu_1^2 +3\alpha'_1\kappa^2_1 +\alpha'_2\kappa^2_2 +\frac{x'_1}{2}v^2_\rho +\frac{y'_1}{2}v^2_R \,, \nonumber \\
&& M^{2\rm Re}_{22} =
-\mu_1^2 +\alpha'_2\kappa^2_1 +3\alpha'_1\kappa^2_2 +\frac{x'_2}{2}v^2_\rho +\frac{y'_2}{2}v^2_R \,, \nonumber \\
&& M^{2\rm Re}_{33} =
+M^2_\rho +\frac{x'_1}{2}\kappa^2_1 +\frac{x'_2}{2}\kappa^2_2
+3\beta' v^2_\rho +\frac{z'}{2}v^2_R \,, \nonumber \\
&& M^{2\rm Re}_{44} =
-\mu_3^2 +\frac{y'_1}{2}\kappa^2_1 +\frac{y'_2}{2}\kappa^2_2
+\frac{z'}{2}v^2_\rho +3\gamma' v^2_R  \,, \nonumber \\
&& M^{2\rm Re}_{12} =
2\alpha'_2 \kappa_1\kappa_2 \,, \nonumber \\
&& M^{2\rm Re}_{13} =
x'_1 \kappa_1v_\rho -\frac{1}{\sqrt2} M'v_R \,, \nonumber \\
&& M^{2\rm Re}_{23} =
x'_2 \kappa_2v_\rho -\frac{1}{\sqrt2} M'v_R \,, \nonumber \\
&& M^{2\rm Re}_{14} =
y'_1 \kappa_1v_R -\frac{1}{\sqrt2} M'v_\rho \,, \nonumber \\
&& M^{2\rm Re}_{24} =
y'_2 \kappa_2v_R -\frac{1}{\sqrt2} M'v_\rho \,, \nonumber \\
&& M^{2\rm Re}_{34} =
z' v_\rho v_R -\frac{1}{\sqrt2} M' (\kappa_1 + \kappa_2) \,,
\end{eqnarray}
while in the basis of $\{ {\rm Im}\phi^0_1,\, {\rm Im}\phi^0_2,\, {\rm Im}\rho^0,\, {\rm Im}\delta^0_R \}$, the elements for the pseudoscalar mass-square matrix are
\begin{eqnarray}
&& M^{2\rm Im}_{11} =
-\mu_1^2 +\alpha'_1\kappa^2_1 +\alpha''_2\kappa^2_2 +\frac{x'_1}{2}v^2_\rho +\frac{y'_1}{2}v^2_R \,, \nonumber \\
&& M^{2\rm Im}_{22} =
-\mu_1^2 +\alpha''_2\kappa^2_1 +\alpha'_1\kappa^2_2 +\frac{x'_2}{2}v^2_\rho +\frac{y'_2}{2}v^2_R \,, \nonumber \\
&& M^{2\rm Im}_{33} =
+M^2_\rho +\frac{x'_1}{2}\kappa^2_1 +\frac{x'_2}{2}\kappa^2_2
+\beta' v^2_\rho +\frac{z'}{2}v^2_R \,, \nonumber \\
&& M^{2\rm Im}_{44} =
-\mu_3^2 +\frac{y'_1}{2}\kappa^2_1 +\frac{y'_2}{2}\kappa^2_2
+\frac{z'}{2}v^2_\rho +\gamma' v^2_R  \,, \nonumber \\
&& M^{2\rm Im}_{12} =
-4 \alpha''_1 \kappa_1\kappa_2 \,, \nonumber \\
&& M^{2\rm Im}_{13} =
+\frac{1}{\sqrt2} M'v_R \,, \nonumber \\
&& M^{2\rm Im}_{23} =
-\frac{1}{\sqrt2} M'v_R \,, \nonumber \\
&& M^{2\rm Im}_{14} =
-\frac{1}{\sqrt2} M'v_\rho \,, \nonumber \\
&& M^{2\rm Im}_{24} =
+\frac{1}{\sqrt2} M'v_\rho \,, \nonumber \\
&& M^{2\rm Im}_{34} =
-\frac{1}{\sqrt2} M' (\kappa_1 + \kappa_2) \,,
\end{eqnarray}
with
\begin{eqnarray}
&& \alpha''_1 = 2\alpha_2 +\alpha_6 \,, \nonumber \\
&& \alpha''_2 = \alpha_1 -4\alpha_2 +2\alpha_3 +\alpha_5 -2\alpha_6 \,.
\end{eqnarray}
When the minimization conditions are applied, the mass-square matrix for the pseudoscalars is greatly simplified,
\begin{eqnarray}
\left( \begin{array}{cccc}
-4 \alpha _1'' \kappa _2^2 + \frac{ M' v_\rho v_R }{\sqrt2 \kappa _1}&
-4 \alpha _1'' \kappa_1 \kappa_2 &
 \frac{ M' v_R }{\sqrt{2}} &
-\frac{ M' v_{\rho } }{\sqrt{2}} \\
-4 \alpha _1'' \kappa_1 \kappa_2 &
-4 \alpha _1'' \kappa _1^2 + \frac{M' v_\rho v_R }{\sqrt2 \kappa_2} &
-\frac{M' v_R}{\sqrt{2}} &
 \frac{M' v_{\rho }}{\sqrt{2}} \\
 \frac{M' v_R}{\sqrt{2}} &
-\frac{M' v_R}{\sqrt{2}} &
 \frac{M' (\kappa_1+\kappa_2) v_R}{\sqrt{2} v_{\rho}} &
-\frac{M' (\kappa_1+\kappa_2) }{\sqrt{2}} \\
-\frac{M' v_{\rho }}{\sqrt{2}} &
 \frac{M' v_{\rho }}{\sqrt{2}} &
-\frac{M' (\kappa _1+\kappa _2)}{\sqrt{2}} &
 \frac{M'  \left(\kappa_1+\kappa_2\right) v_{\rho}}{\sqrt{2} v_R} \\
\end{array} \right) \,.
\end{eqnarray}
As we expect, This matrix has two massless Goldstone boson states, one of which is predominately from $({\rm Im} \delta_R^0)$ and is ``eaten''  by $Z'$ and the other one from combination of the imaginary parts of $\phi_{1,\,2}^0$ and $\rho^0$ becoming the longitudinal component of $Z$. The lighter one of the two massive states, namely $A_{1}$, is expected to lie at the TeV scale, and the other one $A_2$ being a bit heavier, as explicitly shown in Fig.~\ref{scattering-plot}.

At the leading order in the parameters $\frac{\kappa_1,\, \kappa_2,\, v_\rho}{v_R}$, with help of the equations (\ref{parameter-LO}), the mass-square matrix for the CP-even scalars comes out very simple,
\begin{eqnarray}
M^{2{\rm Re}} = {\rm diag}  \{ 0,\, 0,\, 0,\, 2\gamma'v_R^2 \} \,,
\end{eqnarray}
implying that of the four CP-even scalars, only one is at the right-handed scale, and the other three are at lower scales.
When we include the next-to-leading order of $\frac{\kappa_1,\, \kappa_2,\, v_\rho}{v_R}$terms, the matrix becomes
\begin{eqnarray}
M^{2{\rm Re}} = v_R^2 \left( \begin{matrix}
 \frac{1}{\sqrt2} \frac{v_\rho^2}{\kappa_1v_R} &
 0 &
-\frac{1}{\sqrt2} \frac{v_\rho}{v_R} &
 y'_1             \frac{\kappa_1}{v_R} \\
 0 &
 \frac{1}{\sqrt2} \frac{v_\rho^2}{\kappa_2v_R} &
-\frac{1}{\sqrt2} \frac{v_\rho}  {v_R} &
 y'_2             \frac{\kappa_2}{v_R} \\
-\frac{1}{\sqrt2} \frac{v_\rho}{v_R} &
-\frac{1}{\sqrt2} \frac{v_\rho}{v_R} &
 \frac{1}{\sqrt2} \frac{\kappa_1 + \kappa_2}{v_R} &
 z'               \frac{v_\rho}{v_R} \\
 y'_1             \frac{\kappa_1}{v_R} &
 y'_2             \frac{\kappa_2}{v_R} &
 z'               \frac{v_\rho}{v_R} &
 2 \gamma'
\end{matrix} \right),
\end{eqnarray}
in which $y'_1 \kappa_1 = y'_2 \kappa_2$, and we have set $M' = v_\rho$ for simplicity\footnote{There are enough parameters in the potential to make such a choice possible.}. As the right-handed scale is much higher than the electroweak scale $\kappa_1,\, \kappa_2,\, v_\rho$, the masses for three lighter scalars are mainly from the upper-left $3\times3$ block of the matrix (corresponding to the scalars Re$\phi_{1,\,2}^0$ and Re$\phi^0$),
\begin{eqnarray}
v_R^2 \left( \begin{matrix}
 \frac{1}{\sqrt2} \frac{v_\rho^2}{\kappa_1v_R} &
 0 &
-\frac{1}{\sqrt2} \frac{v_\rho}{v_R} \\
 0 &
 \frac{1}{\sqrt2} \frac{v_\rho^2}{\kappa_2v_R} &
-\frac{1}{\sqrt2} \frac{v_\rho}  {v_R} \\
-\frac{1}{\sqrt2} \frac{v_\rho}{v_R} &
-\frac{1}{\sqrt2} \frac{v_\rho}{v_R} &
 \frac{1}{\sqrt2} \frac{\kappa_1 + \kappa_2}{v_R}
\end{matrix} \right) \,.
\end{eqnarray}
One can easily find that one of the eigenstates is massless in the direction of $(\kappa_1,\, \kappa_2,\, v_\rho)$, then there is always a state in the full $4\times4$ matrix with negative values of mass square (this mass eigenstate is expected to play the role of the ``light'' SM Higgs with mass of 125 GeV, after higher order terms are included that makes this negative value positive). 
The two remaining states are expected to be at the scale of $\sqrt{v_{}v_R}$. To get the four mass eigenstates all with positive masses-square values (especially the 125 GeV SM Higgs), we need to expand the mass-square matrix to the next order,
\begin{eqnarray}
M^{2{\rm Re}} = v_R^2 \left( \begin{matrix}
 \frac{1}{\sqrt2} \frac{v_\rho^2}{\kappa_1v_R} &
 \frac{2 \alpha'_2 \kappa_1 \kappa_2}{v_R^2} &
-\frac{1}{\sqrt2} \frac{v_\rho}{v_R} &
 y'_1             \frac{\kappa_1}{v_R} \\
 \frac{2 \alpha'_2 \kappa_1 \kappa_2}{v_R^2} &
 \frac{1}{\sqrt2} \frac{v_\rho^2}{\kappa_2v_R} &
-\frac{1}{\sqrt2} \frac{v_\rho}  {v_R} &
 y'_2             \frac{\kappa_2}{v_R} \\
-\frac{1}{\sqrt2} \frac{v_\rho}{v_R} &
-\frac{1}{\sqrt2} \frac{v_\rho}{v_R} &
 \frac{1}{\sqrt2} \frac{\kappa_1 + \kappa_2}{v_R} &
 z'               \frac{v_\rho}{v_R} \\
 y'_1             \frac{\kappa_1}{v_R} &
 y'_2             \frac{\kappa_2}{v_R} &
 z'               \frac{v_\rho}{v_R} &
 2 \gamma'
\end{matrix} \right) \,.
\end{eqnarray}
The new terms in the (1, 2) (and (2, 1)) element would compensate the negative contribution from mixing with fourth heavy state and then produce a positive value for mass-square of the would-be lightest state (the SM Higgs), thus the quartic coupling combination $\alpha'_2$ is expected to have large value.

\end{document}